\documentclass[letterpaper]{ptephy_v1}

\usepackage{subfig}
\usepackage{amsmath}
\usepackage{graphics}
\usepackage{url}

\usepackage{color}

\def\edth{{\rlap{$\partial$}\raise0.3em\hbox{$-$}}}
\newcommand{\bea}{\begin{eqnarray}}
\newcommand{\eea}{\end{eqnarray}}

\newcommand{\msun}{{\rm M}_{\odot}}
\newcommand{\rsun}{{\rm R}_{\odot}}

\begin{document}

\title{Possible confirmation of the existence of ergoregion
by the Kerr quasinormal mode in gravitational waves from Pop III massive black hole binary}


\author{Tomoya Kinugawa, Hiroyuki Nakano and Takashi Nakamura}

\address{Department of Physics, Kyoto University, Kyoto 606-8502, Japan}

\begin{abstract}
The existence of the ergoregion of the Kerr space-time has not been confirmed 
observationally yet. We show that the confirmation would be possible
by observing the quasinormal mode in gravitational waves. As an example, 
using the recent population synthesis results of Pop III binary black holes,
we find that the peak of the final merger mass ($M_f$) is about $50\msun$,
while the fraction of the final spin $q_f = a_f/M_f > 0.7$ needed
for the confirmation of a part of ergoregion is $\sim 77\%$.
To confirm the frequency of the quasinormal mode, ${\rm SNR} > 35$ is needed.
The standard model of Pop III population synthesis tells us that
the event rate for the confirmation of more than $50\%$ of the ergoregion
by the second generation gravitational wave detectors
is $\sim 2.3$~${\rm events~yr^{-1}~(SFR_p/(10^{-2.5}~M_\odot~yr^{-1}~Mpc^{-3}))}
\cdot (\rm [f_b/(1+f_b)]/0.33)$
where ${\rm SFR_p}$ and ${\rm f_b}$ are the peak value of the Pop III star
formation rate and the fraction of binaries, respectively.
\end{abstract}

\subjectindex{E31, E02, E01, E38}

\maketitle

\section{Introduction}

The Kerr space-time~\cite{Kerr:1963ud} is the unique one in two senses.
Firstly it is the unique stationary
solution~\cite{Israel:1967wq,Carter:1971zc,Robinson:1975bv} of the Einstein equation
in the vacuum under the cosmic censorship~\cite{Penrose:1969pc},
which demands the singularity should be covered by the event horizon.
Secondly, it has the ergoregion where the timelike Killing vector
turns out to be spacelike.
This causes various interesting mechanisms to extract
the rotational energy of the Kerr black hole (BH)
such as Penrose process~\cite{Penrose:1969pc}
and the Blanford-Znajek process~\cite{Blandford:1977ds}.
Although there are so many papers using these two mechanisms
in the fields of physics and astrophysics, 
so far the existence of the ergoregion of the Kerr BH has not been
confirmed observationally yet.
In this paper, we suggest a possible method to confirm
 the existence of  at least a part of the ergoregion.

The Kerr space-time in the geometric unit of $G=c=1$ is specified
by its gravitational mass $M$ and the specific angular momentum $a$.
We use the non-dimensional spin parameter $q=a/M$
instead of $a$ hereafter. There are two important quantities.
The first one is the outer event horizon radius $r_+$ defined by
\bea
 r_{+}=M \left(1 + \sqrt{1-q^2} \right) \,.
\eea
The second one is the location of the outer boundary
of the ergoregion ($r_{\rm ergo}(\theta)$) defined by
\bea
r_{\rm ergo}(\theta) = M \left(1 + \sqrt{1-q^2\cos^2 \theta} \right) \,,
\eea
which is called as the ergosphere.
Note that $r_{\rm ergo}(0)=r_{+}$ and $r_{\rm ergo}(\pi/2)=2M$.

The quasinormal mode (QNM) is the free oscillation of the Kerr BH
after the merger of BH binaries. The complex QNM frequencies are determined
by using the Leaver's method~\cite{Leaver:1985ax} accurately.
A recent numerical relativity simulation of
the BH binary with the initial equal mass
and spins of $q_1=q_2=0.994$ 
results in the final spin $q_f \sim 0.95$~\cite{Scheel:2014ina}.
As for the physical meaning of QNMs, Schutz and Will~\cite{Schutz:1985zz}
used the WKB method for the $q_f=0$ case, that is, the Schwarzschild space-time, 
and they showed that the real and imaginary parts of the QNMs
are determined by the peak value and the second derivative , respectively,
of  the Regge-Wheeler potential~\cite{Regge:1957td},
which determines the behavior of gravitational perturbations 
in the Schwarzschild space-time.
The location of the peak for the dominant $\ell=2$ mode is
at $r_{\rm max}=3.28M$, and the errors due to the WKB approximation
are about $7\%$ and $0.7\%$
for the real and imaginary parts of the fundamental ($n=0$) QNM frequency,
respectively.
This suggests that the complex frequency of the QNM is determined
by the space-time around $r_{\rm max}$.
Conversely, if the $\ell=2$ QNM which is the dominant mode, 
is confirmed by the second generation gravitational wave detectors,
such as Advanced LIGO (aLIGO)~\cite{TheLIGOScientific:2014jea}, 
Advanced Virgo (AdV)~\cite{TheVirgo:2014hva}, 
and KAGRA~\cite{Somiya:2011np,Aso:2013eba},
we can say that the strong space-time around $r=3.28M$ is confirmed
as predicted by Einstein's general relativity.
The reason for the word ``around'' comes from the fact that the imaginary part 
of the QNM frequency is determined by the second derivative  of the Regge-Wheeler potential.

In the  paper submitted to PTEP~\cite{Nakano:2016a} by Nakano, Nakamura and Tanaka ,
they  showed that the similar physical picture to that presented by Schutz and Will
can be obtained by using the Detweiler potential~\cite{Detweiler:1977gy}
of gravitational perturbations~\cite{Teukolsky:1973ha} 
in the Kerr space-time. \footnote{Note that the Detweiler potential  corresponds to either the Regge-Wheeler
or Zerilli~\cite{Zerilli:wd} one in the Schwarzschild space-time
(see also Ref.~\cite{Nakamura:2016gri}).}
The maximum errors of the real and imaginary parts of the QNM frequency with ($\ell=m=2$),
which is the dominant mode shown by the numerical relativity simulations~\cite{London:2014cma},
are $\lesssim 1.5\%$ and $\lesssim 2\%$ in the range of $0.7 < q_f < 0.98$, respectively.
They  also obtained that the QNM for $q_f  > 0.7$ reflects the Kerr space-time
within the ergoregion because $r_{\rm max} < 2M$.
Since the ergoregion radius depends on $\theta$,
we can define the covered solid angle $4\pi C$ for each $r_{\rm max} < 2M$
by $C=\cos \theta_m$ with  $\theta_m$ defined
by $r_{\rm max} =M (1 + \sqrt{1-q^2\cos^2 \theta_m} )$.
It is found that an empirical relation between $(1-C)$ and $(1-q)$ for $0.7 < q < 0.98$
 given by
\begin{equation}
 \ln(1-C) =2.7867 \,\ln(1-q) + 3.0479 \,,
\label{eq:empiricalC}
\end{equation}
exists.
The correlation coefficient of this empiriacl relation is $0.989$
with the chance probability of $5.9\times 10^{-8}$.

The purpose of this paper is to apply Eq.~\eqref{eq:empiricalC}
to the recent population synthesis results of  Population III (Pop III) massive BH binaries 
to know the event rate of detection of the QNM gravitational waves
and the typical value of $C$ as an example.

This paper is organized as follows.
In \S 2, we briefly argue
the recent population synthesis results of Pop III massive BH
binaries~\cite{Kinugawa:2014zha,Kinugawa:2015nla}.
The reader who is not familiar with the population synthesis, may skip this section.
In \S 3, we discuss methods to obtain the final BH's mass $M_f$
and spin $q_f$ with their distribution functions and
the detection rate as a function of $C$.
Finally, \S 4 is devoted the discussions.

\section{Pop III binary calculation}
PopIII star is the first star in our universe which does not have the metal with atomic number
larger than the carbon.
To study Pop III binary evolutions, they used  a Pop III binary population
synthesis code~\cite{Kinugawa:2014zha,Kinugawa:2015nla}
which is upgraded from Hurley's BSE code~\cite{Hurley:2002rf,hurleysite}
for the case of Pop I stars to that of Pop III stars.
They  calculated $10^6$ binary evolutions
for  given initial values of the primary mass $M_1$,
the mass ratio $M_2/M_1$,
the orbital separation $a$ and the eccentricity $e$
using the  Monte Carlo method under the initial distribution functions.
They call  the primary star as  the larger mass one while the secondary is the smaller mass one in the binary.
The typical mass of Pop III stars is from $\sim 10\,\msun$
to $\sim 100\,\msun$~\cite{Hosokawa:2011qa,Hosokawa:2012pp}.
Thus, they  took  the initial mass function which may be
flat from $10\,\msun$ to $140\,\msun$ suggested from the numerical simulations~\cite{Hirano:2013lba, Susa:2014moa}.
The reason for  the upper limit of mass of  $140\,\msun$ is that
the star with mass larger than $140\,\msun$ becomes the pair instability supernova leaving  no remnant.
Since there is no observation of Pop III stars and binaries,
they  simply assume that other initial distribution functions
are the same as Pop I binaries.
The initial mass ratio function for given $M_1$ is flat from $10\,\msun/M_1$ to $1$.
The separation $a$ \footnote{This ``$a$'' is completely different from ``$a$'' 
in the Kerr BH so that we use $q=a/M$ as the Kerr parameter.} distribution function is proportional
to $1/a$ from $a_{\rm min}$ to $10^6\,\rsun$,
where $a_{\rm min}$ is the minimum separation
when the binary interaction such as the mass transfer and so on is absent.
The initial eccentricity distribution function
is proportional to $e$ from $0$ to $1$.
The set of these initial distribution functions
is the same as their standard model with $140$ case of Ref.~\cite{Kinugawa:2015nla}.
In this paper, we choose the binary evolution parameters of their standard model
and the optimistic core-merger criterion of Ref.~\cite{Kinugawa:2015nla}.
The details of binary interactions and spin evolution, which is very important in this paper,
are discussed in Refs.~\cite{Kinugawa:2014zha, Kinugawa:2015nla}.  

\begin{figure}[!ht]
\begin{center}
\includegraphics[width=0.62\textwidth,clip=true]{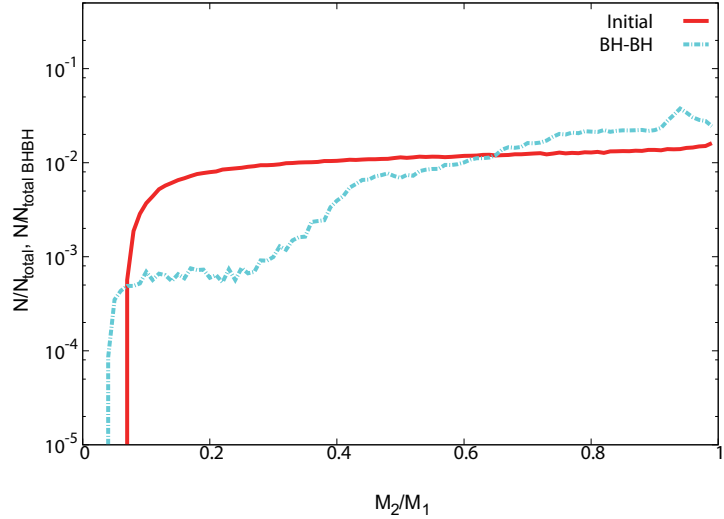}
\end{center}
 \caption{The distribution of mass ratio $M_2/M_1\leq 1$.
 The distributions of the initial mass ratio  and the one
 when the binaries become BH-BHs  are shown as red and light blue lines, respectively.
 The initial mass ratio distribution is normalized
 by the total binary number $N_{\rm total}=10^6$ while 
 the one  when the binaries become BH-BHs
 is normalized by the total binary number $N_{\rm total \,BHBH}=128897$.}
\label{fig:massratio}
\end{figure}

In Ref.~\cite{Kinugawa:2015nla}, they found that 
$\sim 13\,\%$ of Pop III binaries become BH-BH binaries 
which merge within the Hubble time and the typical mass of Pop III BHs
is $\sim 30\,\msun$.
Figure~\ref{fig:massratio} shows the initial mass ratio distribution
of Pop III binaries (red line) and that of
Pop III BH-BHs (blue dashed line).
Even though the mass ratio of binaries smaller than $\sim 0.5$ exists substantially  initially,
most of BH-BH  binaries  has mass ratio larger than $\sim 0.5$ by the effect of mass transfer.
Thus, large mass ratio ($=M_2/M_1$) BH-BHs are the majority.

\begin{figure}[!ht]
\begin{center}
\includegraphics[width=0.62\textwidth,clip=true]{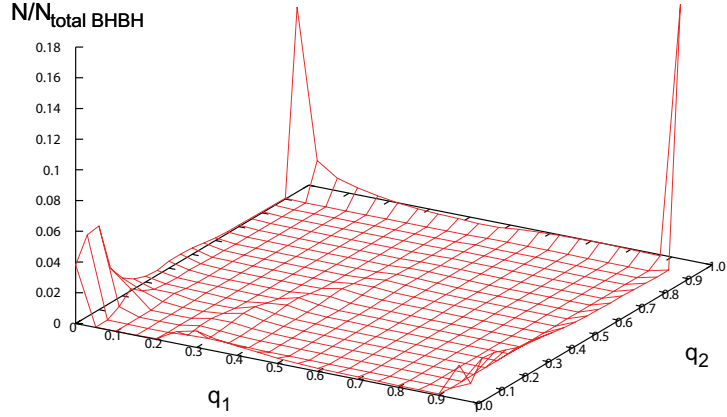}
\end{center}
 \caption{The distribution of spin parameters.
 The  distribution of spin parameters
 when each star becomes BH is shown.
 $q_1$ and $q_2$ are the spin parameters of the primary and the secondary BHs, respectively.
 This distribution when the binaries become BH-BHs is normalized
 by the total binary number $N_{\rm total \,BHBH}=128897$ with the grid separation being $\Delta q_1 = \Delta q_2=0.05$.}
 \label{fig:spin1spin2}
\end{figure}

\begin{figure}[t]
    \begin{center}
        \subfloat[$0<q_1<0.05$]{
            \includegraphics[scale=0.5]{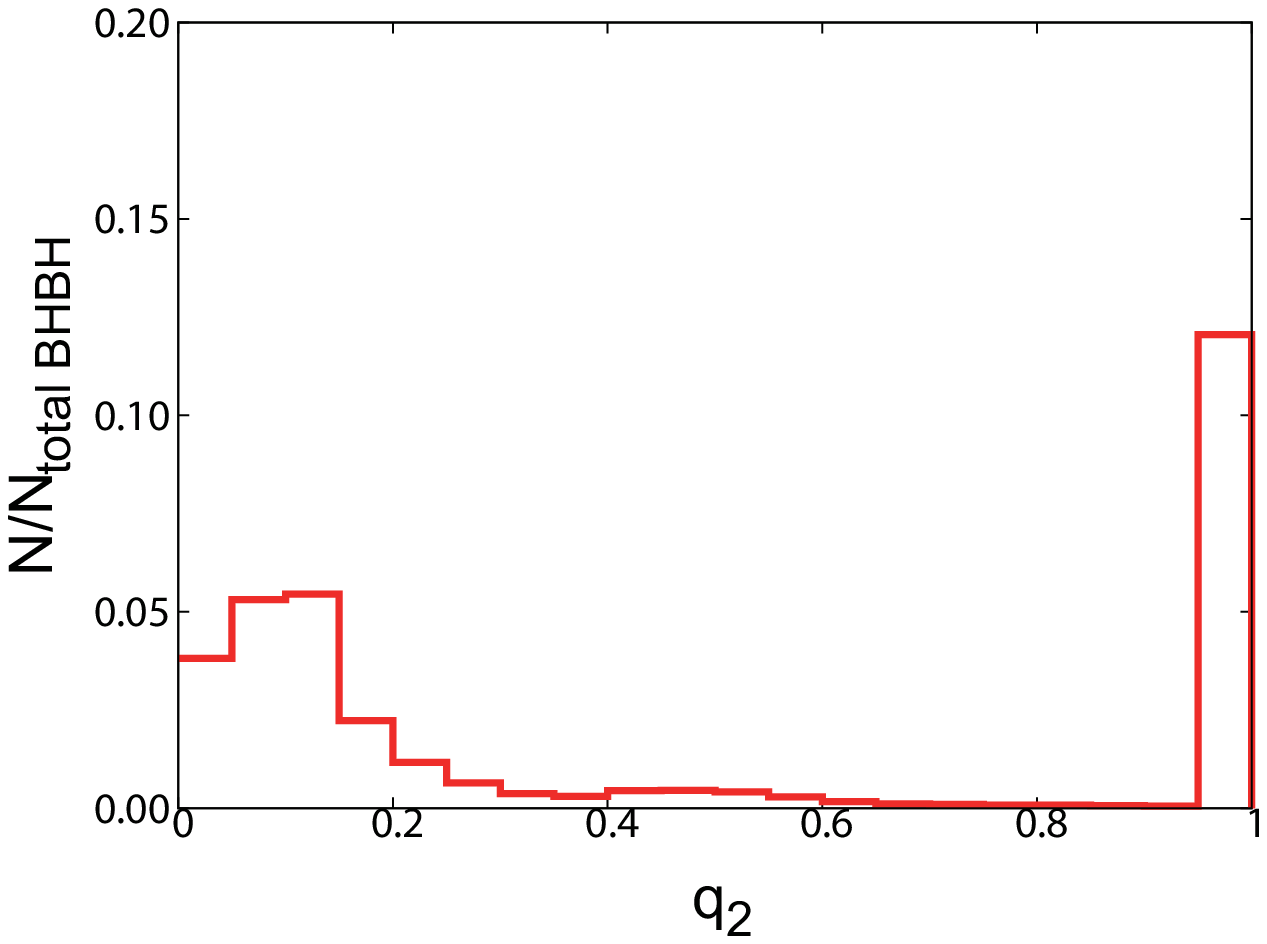}
        }
        \subfloat[$0.95<q_1<0.998$]{
            \includegraphics[scale=0.5]{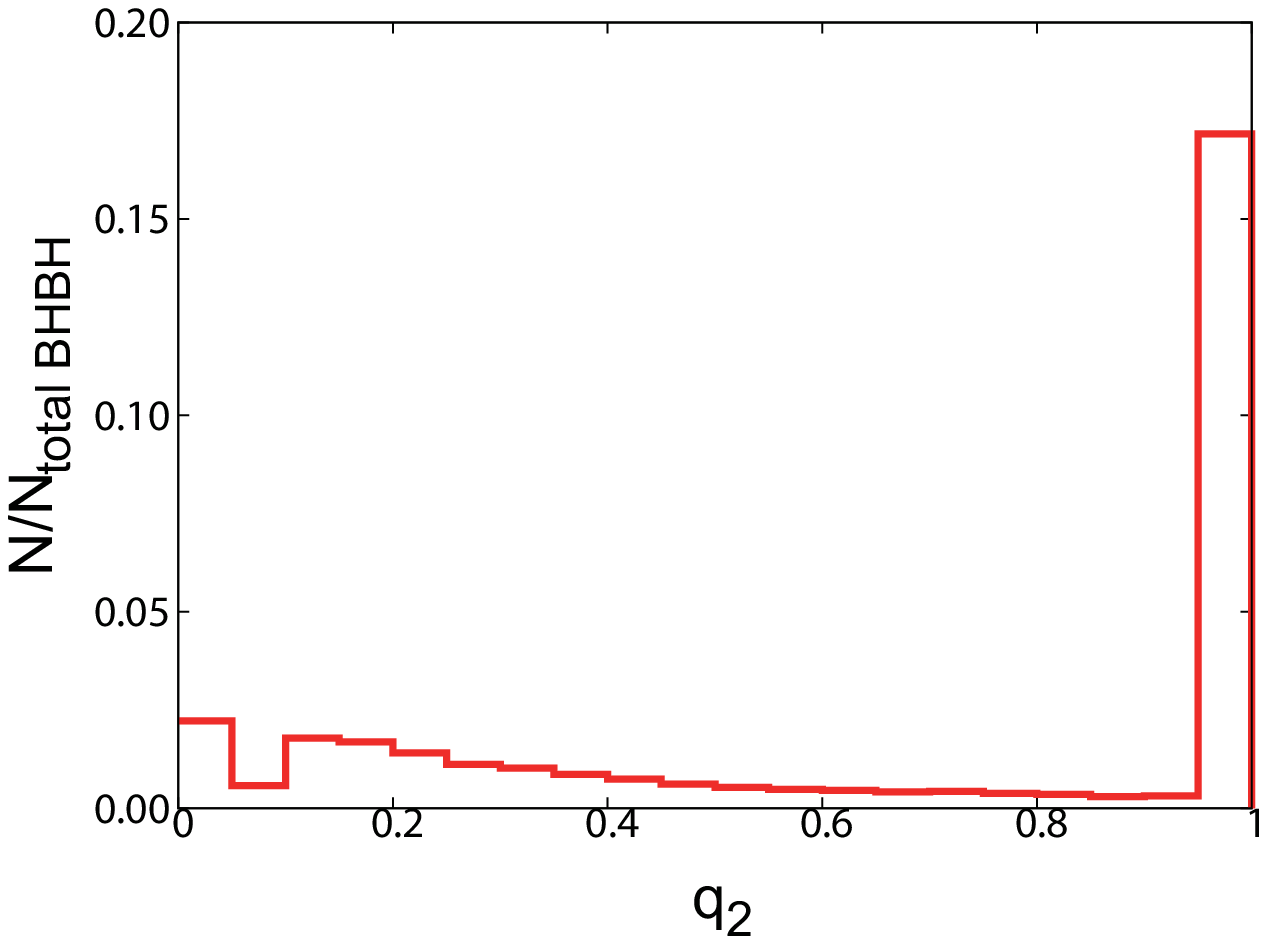}
        }
    \end{center}    
\caption{Cross section views of distributribution of spin parameter.}
(a) The distribution of $q_2$ for $0 < q_1 < 0.05$.  We can see  that $q_2$ distribution has bimodial peaks at  $0<q_2<0.15$ and $0.95<q_2<0.998$ .
(b) The distribution of $q_2$ for $0.95 < q_1 < 0.998$. We see that the large value of $q_2$ is the majority
so that  there is a group in which both $q_1$ and $q_2$ are large.
\label{fig:cross section}
\end{figure}

Figure~\ref{fig:spin1spin2} shows  the distribution
of spin parameters of the primary and secondary BHs
when the primary and secondary become BHs.
Figure~\ref{fig:cross section} shows the cross section views of distributribution of spin parameter with
(a) the cross section views of distributribution of spin parameter when $0<q_1<0.05$ and 
(b) the cross section views of distributribution of spin parameter when $0.95<q_1<0.998$.
The spin parameter of each BH is calculated
by the angular momentum of the progenitor just before it becomes BH.
If the spin parameter of the BH is larger than the Thorne limit~\cite{Thorne:1974ve},
we assign $q=q_{\rm Thorne}=0.998$ as  the spin parameter.
From Fig.~\ref{fig:spin1spin2} and Fig.~\ref{fig:cross section}, the spin parameters of Pop III BH-BHs are roughly classified
into 3 groups.
First, the majority of Pop III is in the group 
where both BHs have high spin parameters.
If the mass transfer is dynamically unstable
or the secondary plunges into the primary envelope,
the orbit shrinks and the primary envelope is stripped by the friction
between the secondary and the primary envelope~\cite{Webbink:1984ti}.
In this group, the progenitors evolve without the common envelope phase
and the primary envelope is not stripped.
Thus, BHs of this group get large angular momentum from the envelope of progenitor
and the spin parameter which has the largest Thorne limit $q_{\rm Thorne}=0.998$.
Second, there is a group where both BHs have low spin parameters.
In this group, each star evolves via the common envelope phase
and they take off their envelope and lose almost all of the angular momentum.
Thus, there are many Pop III BH-BHs with $q_1<0.15,\,q_2<0.15$.
Third, there is a group where the one of the pair has high spin
and the other has low spin.
In this group, the primary evolves with the common envelope phase
and the secondary evolves without the common envelope, or vice versa. 


\section{Remnant mass, spin and the detection rate}

Given BH binary parameters, $M_1$, $M_2$, $q_1$ and $q_2$,
we calculate the remnant mass and spin by using formulae from spin aligned
BH binaries~\cite{Lousto:2009mf, Healy:2014yta}
(see also Ref.~\cite{Barausse:2009uz, Barausse:2012qz} from a different group).

The final (non-dimensional) spin parameter $q_f$ is
\bea
q_f
= \frac{S_f}{M_f^2}
= (4\eta)^2 \left( L_0 + L_{1}\,\tilde S_\| + L_{2a}\,\tilde \Delta_\| \delta m
+ \cdot \cdot \cdot \right)
+ (1+8\eta) \tilde S_\| \delta m^4+\eta\tilde{J}_{\rm ISCO} \delta m^6 \,.
\label{eq:remnantq}
\eea
where
\bea
&& \eta = \frac{M_1M_2}{M^2} \,, \quad  M=M_1+M_2 \,,
\quad \delta m=\frac{M_1-M_2}{M} \,,
\cr
&& \tilde S_\| = \frac{M_1^2 q_1 + M_2^2 q_2}{M^2} \,, \quad
\tilde \Delta_\| = \frac{M_2 q_2 - M_1 q_1}{M} \,,
\eea
and ($+ \cdot\cdot\cdot$) denotes the higher order correction
with respect to spins
which is given in Eq.~(14) of Ref.~\cite{Healy:2014yta} explicitly.
$L_0$, $L_{1}$ and $L_{2a}$ are the fitting parameters summarized in Table VI
of Ref.~\cite{Healy:2014yta}, and the last two terms in Eq.~\eqref{eq:remnantq}
are added to enforce the particle limit ($\eta \to 0$)
where $\tilde{J}_{\rm ISCO}$ is the orbital angular momentum
of the innermost stable circular orbit (ISCO).
The final mass $M_f$ is given by
\bea
\frac{M_f}{M}
= (4\eta)^2 \left( M_0 + K_1 \tilde S_\|+ K_{2a}\,\tilde \Delta_\| \delta m
+ \cdot \cdot \cdot \right)
+\left[1+ \eta(\tilde{E}_{\rm ISCO} + 11)\right] \delta m^6 \,.
\label{eq:remnantM}
\eea
Again, $M_0$, $K_{1}$ and $K_{2a}$ are the fitting parameters summarized in Table VI
of Ref.~\cite{Healy:2014yta}.
In practice, we use Ref.~\cite{Ori:2000zn}
for the ISCO angular momentum and energy,
$\tilde{J}_{\rm ISCO}$ and $\tilde{E}_{\rm ISCO}$
(note that we assign $q_f$ to $a$ in Ref.~\cite{Ori:2000zn}).

According to a recent numerical relativity simulation
for a highly spinning BH binary merger
with $M_1=M_2=1/2$, $q_1=q_2=0.994$~\cite{Scheel:2014ina},
the final mass and spin after merger are obtained as
$M_{\rm f} = 0.887$ and $q_{\rm f} = 0.950$, respectively.
On the other hand, the remnant formulae in Eqs.~\eqref{eq:remnantq}
and \eqref{eq:remnantM}
which are not calibrated by the above numerical relativity result,
give $M_f = 0.888$ and $q_f  = 0.950$.
We see that the formulae are sufficiently accurate for our analysis
(see also a recent study on remnant BHs for precessing BH binaries~\cite{Zlochower:2015wga}).
The radiated energy is so large that
the total mass of $60\,\msun$ for the above highly spinning binaries
becomes the remnant mass of $53.28\,\msun$.
We note that Eq.~\eqref{eq:remnantq} cannot give any realistic solution
for some large mass ratio, e.g., 
for $\eta \gtrsim 0.1249$ ($q_1=q_2=0.998$), 
$\eta \gtrsim 0.1169$ ($q_1=q_2=0.994$) 
and $\eta \gtrsim 0.1066$ ($q_1=q_2=0.99$). 
In that case, we simply set $q_f  = 0.998$.

\begin{figure}[!ht]
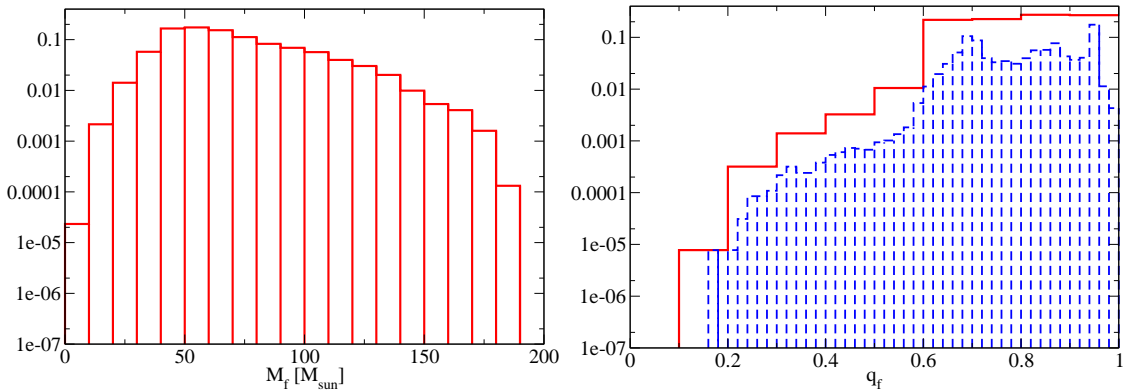

\begin{center}
 \includegraphics[width=0.48\textwidth,clip=true]{./Mrem}
 \includegraphics[width=0.48\textwidth,clip=true]{./qrem}
\end{center}
 \caption{(Left) The normalized distribution of $M_f$ obtained by binning
 with $\Delta M_f=10\,\msun$.
 (Right) The normalized distribution of $q_f$. The solid red and dashed blue lines are
 obtained by binning with $\Delta q_f=0.1$ and $0.02$, respectively.}
 \label{fig:remnant}
\end{figure}

In Fig.~\ref{fig:remnant}, we show the remnant mass and spin
calculated by the remnant formulae.
Due to the mass decrease by the gravitational wave radiation,
we see the peak in a bin between $50\,\msun$ and $60\,\msun$
which reflects the peak of the total mass of Pop III BH-BHs.
The remnant spin $q_f>0.96$ is $1.56\%$,
and only $0.429\%$ of the remnant BHs have the spin
larger than $0.98$.

To estimate the signal-to-noise ratio (SNR)
of the QNM (ringdown) signal
in the expected noise curve of KAGRA~\cite{Somiya:2011np,Aso:2013eba} 
[bKAGRA, VRSE(D) configuration] shown in Ref.~\cite{bKAGRA},
we use the results derived by Flanagan and Hughes in Ref.~\cite{Flanagan:1997sx}.
In Ref.~\cite{Nakano:2015uja}, we have fitted the KAGRA noise curve as
\bea
S_n(f)^{1/2} = 10^{-26} \left( 6.5\times 10^{10} f^{-8}
+ 6\times 10^6 f^{-2.3}
+ 1.5 f^1 \right) \, [{\rm Hz^{-1/2}}] \,,
\eea
where the frequency $f$ is in units of Hz.
According to Ref.~\cite{Flanagan:1997sx},
the angle averaged SNR for the ringdown phase is calculated from
Eq.~(B14) of Ref.~\cite{Flanagan:1997sx} as
\bea
{\rm SNR} =
\sqrt{\frac{128}{5}}
\frac{\eta}{F(q_f)} \sqrt{\frac{\epsilon_r M_f}{S_n(f_c)}}\frac{M_f}{D} \,,
\label{eq:SNR}
\eea
where $F(q_f)$ and $f_c$ are given in Ref.~\cite{Berti:2005ys},
\bea
F(q_f) = 1.5251 - 1.1568 (1-q_f)^{0.1292} \,,
\quad
f_c = \frac{1}{2\pi M_{f}} F(q_f) \,.
\eea
and $\epsilon_r$ denotes the fraction of the total mass energy radiated
in the ringdown phase which is assumed as $\epsilon_r=0.03$.
Here, we have ignored effects of the redshifted mass and the cosmological distance,
i.e., the redshift and the difference between the distance $D$
and the luminosity distance.
Since the maximum distance considered here is $z \sim 0.28$, the errors are small.
Although the calculation is straightforward,
we do not show the explicit expression
since the expression is complicated due to $S_n(f_c)$.
For example, we have ${\rm SNR}=23$
in the case of $M_f=60\,\msun$, $q_f=0.7$, $\eta=1/4$ and $D=200$~Mpc.

\begin{figure}[!ht]
\begin{center}
 \includegraphics[width=0.6\textwidth,clip=true]{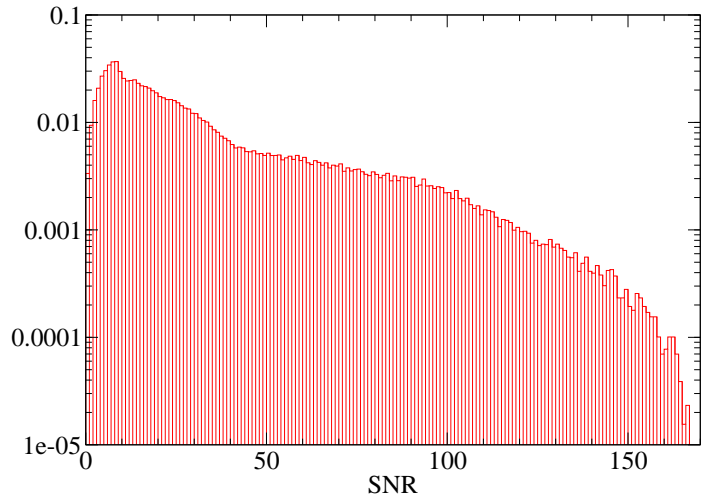}
\end{center}
 \caption{The normalized distribution of the SNRs obtained by binning
 with $\Delta {\rm SNR}=1$ calculated from Eq.~\eqref{eq:SNR} for $D=200$~Mpc.}
 \label{fig:SNR_200Mpc}
\end{figure}

Figure~\ref{fig:SNR_200Mpc} shows the normalized distribution of the SNRs.
Here, we have assumed all gravitational wave sources are located at $D=200$~Mpc.
To calculate the detection rate, we need to know the merger rate density
of Pop III BH-BHs. The merger rate density
derived in Ref.~\cite{Kinugawa:2015nla}
is approximated by $R_m = 0.024+0.0080~(D/1\,{\rm Gpc})$~$[{\rm Myr^{-1} Mpc^{-3}}]$
for a low redshift (note that this fitting works up to $z \sim 2$).

In this paper, we focus on the detection rate
of the solid angle of a sphere emitting the QNM, $4\pi C$
which dips in the ergoregion. 
In Ref.~\cite{Nakano:2016a}, we obtained a simple relation between $C$
and the spin parameter $q$ shown as Eq.~\eqref{eq:empiricalC} in Introduction.
In Table~\ref{tab:detection_C}, we present the result for the detection rate
in the case of ${\rm SNR}=35$ which is needed to confirm the QNM~\cite{Berti:2007zu}.

\begin{table}[!ht]
\caption{
 The detection rate $[{\rm yr}^{-1}]$ devided by dependence on the star formation rate ${\rm SFR_p}$ and the fraction of the binary ${\rm f_b}$ as a function  of the lower limit of  the solid angle of a sphere $4\pi C$
where the QNM is  mainly emitted from the ergoregion 
 in the case of ${\rm SNR}=35$ for the KAGRA detector.
}
\label{tab:detection_C}
\begin{center}
\begin{tabular}{|c|c|c|c|c|c|c|}
\hline
$0<C$ & $0.5<C$ & $0.7<C$ & $0.9<C$ & $0.95<C$ & $0.97<C$ & $0.99<C$ \\
\hline 
3.73 & 2.23 & 1.10 & 0.356 & 0.162 & 0.117 & 0.0780
\\
\hline
\end{tabular}
\end{center}
\end{table}

\section{Discussion}
In this paper, we argued only BH-BH binary from Pop III star origin as an example. 
In Table~\ref{tab:detection_C},
the standard model of Pop III population
synthesis~\cite{Kinugawa:2014zha, Kinugawa:2015nla} tells us that
the event rate for the confirmation of more than $50\%$ of the ergoregion
by the second generation gravitational wave detectors
is $\sim 2.3$~${\rm events~yr^{-1}~(SFR_p/(10^{-2.5}~M_\odot~yr^{-1}~Mpc^{-3}))}
\cdot (\rm [f_b/(1+f_b)]/0.33)$
where ${\rm SFR_p}$ and ${\rm f_b}$ are the peak value of the Pop III star
formation rate and the fraction of binaries, respectively.
Here, we set ${\rm SNR} = 35$ because at least
this SNR is needed to confirm the QNM frequency.
Furthermore, by massive Pop I and Pop II binaries,
the above rate could get larger.
Here Pop I stars is the sun like one with $\sim 2\%$ metal in the weight
while Pop II is the old star like in the globular cluster with $\sim 10^{-2}\%$ metal. 
The Pop I and Pop II binary population
synthesis~\cite{Dominik:2012, Dominik:2013, Dominik:2014}
showed that the some fraction of Pop II evolves into massive BH-BHs.
In addition, the rotating Pop II stars are easier to become massive BH-BHs
than non-rotating Pop II stars~\cite{de Mink:2016, Marchant:2016}.
Since the mass 
loss is expected for Pop I and Pop II stars due to the absorption of photons at the spectral 
lines of metals, these Pop I and Pop II cases, however, the progenitors are easier
to lose the angular momentum by the stellar wind mass loss and so on.
Thus, the Pop I and Pop II cases might have lower spin than the Pop III case.
If , however, the mass and the spin distributions
of references  \cite{Dominik:2014,de Mink:2016, Marchant:2016} are available, we can compute the detection rate
of the QNM and the covered solid angle $4\pi C$ of the ergoregion. In this sense, the rate shown in this paper is the minimum one. 

It is noted that from the standard model of Pop III population synthesis,
the event rate of the final $q_f > 0.98$ BHs is
$\sim 5.17 \times 10^{-6}$~${\rm events~yr^{-1}~(SFR_p/(10^{-2.5}~M_\odot~yr^{-1}~Mpc^{-3}))}
\cdot (\rm [f_b/(1+f_b)]/0.33)$ for ${\rm SNR}=35$.
We expect interesting physics in highly/extremely spinning Kerr BHs,
which needs further studies.
To detect such a BH, third generation gravitational wave detectors,
such as the Einstein Telescope (ET)~\cite{Punturo:2010zz}
should be required.
Also, although there is room for highly spinning remnant BHs
with $q_f > 0.98$ from the merger of comparable mass BH binaries,
we will need large-mass-ratio binaries
for which the cosmic censorship~\cite{Penrose:1969pc}
has been discussed extensively
(see e.g., Ref.~\cite{Barausse:2011vx} and references therein).
These binaries could be  one of the targets
for space based gravitational wave detectors
such as eLISA~\cite{Seoane:2013qna} and DECIGO~\cite{Seto:2001qf}
 at the formation time of $z\sim 10$.

\section*{Acknowledgment}

~~This work was supported by MEXT Grant-in-Aid for Scientific Research
on Innovative Areas,
``New Developments in Astrophysics Through Multi-Messenger Observations
of Gravitational Wave Sources'', No.~24103006 (TN, HN) and
by the Grant-in-Aid from the Ministry of Education, Culture, Sports,
Science and Technology (MEXT) of Japan No.~ 251284 (TK) and~15H02087 (TN).




\begin{thebibliography}{99}

\bibitem{Kerr:1963ud} 
  R.~P.~Kerr,
  Phys.\ Rev.\ Lett.\  {\bf 11}, 237 (1963).

\bibitem{Israel:1967wq} 
  W.~Israel,
  Phys.\ Rev.\  {\bf 164}, 1776 (1967).

\bibitem{Carter:1971zc} 
  B.~Carter,
  Phys.\ Rev.\ Lett.\  {\bf 26}, 331 (1971).

\bibitem{Robinson:1975bv} 
  D.~C.~Robinson,
  Phys.\ Rev.\ Lett.\  {\bf 34}, 905 (1975).

\bibitem{Penrose:1969pc} 
  R.~Penrose,
  Riv.\ Nuovo Cim.\  {\bf 1}, 252 (1969)
  [Gen.\ Rel.\ Grav.\  {\bf 34}, 1141 (2002)].

\bibitem{Blandford:1977ds} 
  R.~D.~Blandford and R.~L.~Znajek,
  Mon.\ Not.\ Roy.\ Astron.\ Soc.\  {\bf 179}, 433 (1977).

\bibitem{Leaver:1985ax} 
  E.~W.~Leaver,
  Proc.\ Roy.\ Soc.\ Lond.\ A {\bf 402}, 285 (1985).

\bibitem{Scheel:2014ina} 
  M.~A.~Scheel, M.~Giesler, D.~A.~Hemberger, G.~Lovelace, K.~Kuper, M.~Boyle, B.~Szilagyi and L.~E.~Kidder,
  Class.\ Quant.\ Grav.\  {\bf 32}, 105009 (2015)
  [arXiv:1412.1803 [gr-qc]].

\bibitem{Schutz:1985zz} 
  B.~F.~Schutz and C.~M.~Will,
  Astrophys.\ J.\  {\bf 291}, L33 (1985).

\bibitem{Regge:1957td}
  T.~Regge and J. A. Wheeler,
  Phys.\ Rev.\  {\bf 108}, 1063 (1957).

\bibitem{TheLIGOScientific:2014jea} 
  J.~Aasi {\it et al.}  [LIGO Scientific Collaboration],
  Class.\ Quant.\ Grav.\  {\bf 32}, 074001 (2015)
  [arXiv:1411.4547 [gr-qc]].

\bibitem{TheVirgo:2014hva} 
  F.~Acernese {\it et al.}  [VIRGO Collaboration],
  Class.\ Quant.\ Grav.\  {\bf 32}, 024001 (2015)
  [arXiv:1408.3978 [gr-qc]].

\bibitem{Somiya:2011np} 
  K.~Somiya [KAGRA Collaboration],
  Class.\ Quant.\ Grav.\  {\bf 29}, 124007 (2012)
  [arXiv:1111.7185 [gr-qc]].

\bibitem{Aso:2013eba} 
  Y.~Aso {\it et al.}  [KAGRA Collaboration],
  Phys.\ Rev.\ D {\bf 88}, 043007 (2013)
  [arXiv:1306.6747 [gr-qc]].

\bibitem{Nakano:2016a} 
  H.~Nakano, T.~Nakamura and T.~Tanaka,
  ``The detection of quasinormal mode with $a/M \sim 0.95$ would
prove a sphere $99\%$ soaking in the ergoregion of the Kerr space-time'',
  submitted to Progress of Theoretical and Experimental Physics (PTEP) (2016).
   
\bibitem {Detweiler:1977gy} 
  S.~L.~Detweiler,
  Proc.\ Roy.\ Soc.\ Lond\ A {\bf 352}, 381 (1977).

\bibitem{Teukolsky:1973ha} 
  S.~A.~Teukolsky,
  Astrophys.\ J.\  {\bf 185}, 635 (1973).

\bibitem{Zerilli:wd}
  F.~J.~Zerilli,
  Phys.\ Rev.\ D {\bf 2}, 2141 (1970).

\bibitem{Nakamura:2016gri} 
  T.~Nakamura, H.~Nakano and T.~Tanaka,
Phys.\ Rev.\ D in press
  arXiv:1601.00356 [astro-ph.HE].

\bibitem{London:2014cma} 
  L.~London, D.~Shoemaker and J.~Healy,
  Phys.\ Rev.\ D {\bf 90}, 124032 (2014)
  [arXiv:1404.3197 [gr-qc]].

\bibitem{Kinugawa:2014zha} 
  T.~Kinugawa, K.~Inayoshi, K.~Hotokezaka, D.~Nakauchi and T.~Nakamura,
  Mon.\ Not.\ Roy.\ Astron.\ Soc.\  {\bf 442}, 2963 (2014)
  [arXiv:1402.6672 [astro-ph.HE]].

\bibitem{Kinugawa:2015nla} 
  T.~Kinugawa, A.~Miyamoto, N.~Kanda and T.~Nakamura,
  arXiv:1505.06962 [astro-ph.SR].

\bibitem{Hurley:2002rf} 
  J.~R.~Hurley, C.~A.~Tout and O.~R.~Pols,
  Mon.\ Not.\ Roy.\ Astron.\ Soc.\  {\bf 329}, 897 (2002)
  [astro-ph/0201220].

\bibitem{hurleysite}
 \url{http://astronomy.swin.edu.au/~jhurley/}

\bibitem{Hosokawa:2011qa} 
  T.~Hosokawa, K.~Omukai, N.~Yoshida and H.~W.~Yorke,
  Science {\bf 334}, 1250 (2011)
  [arXiv:1111.3649 [astro-ph.CO]].

\bibitem{Hosokawa:2012pp} 
  T.~Hosokawa, N.~Yoshida, K.~Omukai and H.~W.~Yorke,
  Astrophys.\ J.\  {\bf 760}, L37 (2012)
  [arXiv:1210.3035 [astro-ph.CO]].

\bibitem{Hirano:2013lba} 
  S.~Hirano, T.~Hosokawa, N.~Yoshida, H.~Umeda, K.~Omukai, G.~Chiaki and H.~W.~Yorke,
  Astrophys.\ J.\  {\bf 781}, 60 (2014)
  [arXiv:1308.4456 [astro-ph.CO]].

\bibitem{Susa:2014moa} 
  H.~Susa, K.~Hasegawa and N.~Tominaga,
  Astrophys.\ J.\  {\bf 792}, no. 1, 32 (2014)
  [arXiv:1407.1374 [astro-ph.GA]].

\bibitem{Thorne:1974ve} 
  K.~S.~Thorne,
  Astrophys.\ J.\  {\bf 191}, 507 (1974).

\bibitem{Webbink:1984ti} 
  R.~F.~Webbink,
  Astrophys.\ J.\  {\bf 277}, 355 (1984).

\bibitem{Lousto:2009mf}
 C.~O.~Lousto, M.~Campanelli, Y.~Zlochower and H.~Nakano,
 Class.\ Quant.\ Grav.\ {\bf 27}, 114006 (2010)
 [arXiv:0904.3541 [gr-qc]].

\bibitem{Healy:2014yta} 
  J.~Healy, C.~O.~Lousto and Y.~Zlochower,
  Phys.\ Rev.\ D {\bf 90}, 104004 (2014)
  [arXiv:1406.7295 [gr-qc]].


\bibitem{Barausse:2009uz} 
  E.~Barausse and L.~Rezzolla,
  Astrophys.\ J.\  {\bf 704}, L40 (2009)
  [arXiv:0904.2577 [gr-qc]].

\bibitem{Barausse:2012qz} 
  E.~Barausse, V.~Morozova and L.~Rezzolla,
  Astrophys.\ J.\  {\bf 758}, 63 (2012)
  [Astrophys.\ J.\  {\bf 786}, 76 (2014)]
  [arXiv:1206.3803 [gr-qc]].

\bibitem{Ori:2000zn} 
  A.~Ori and K.~S.~Thorne,
  Phys.\ Rev.\ D {\bf 62}, 124022 (2000)
  [gr-qc/0003032].

\bibitem{Zlochower:2015wga} 
  Y.~Zlochower and C.~O.~Lousto,
  Phys.\ Rev.\ D {\bf 92}, 024022 (2015)
  [arXiv:1503.07536 [gr-qc]].

\bibitem{bKAGRA}
  \url{http://gwcenter.icrr.u-tokyo.ac.jp/researcher/parameters}

\bibitem{Flanagan:1997sx} 
  E.~E.~Flanagan and S.~A.~Hughes,
  Phys.\ Rev.\ D {\bf 57}, 4535 (1998)
  [gr-qc/9701039].

\bibitem{Nakano:2015uja} 
  H.~Nakano, T.~Tanaka and T.~Nakamura,
  Phys.\ Rev.\ D {\bf 92}, 064003 (2015)
  [arXiv:1506.00560 [astro-ph.HE]].

\bibitem{Berti:2005ys} 
  E.~Berti, V.~Cardoso and C.~M.~Will,
  Phys.\ Rev.\ D {\bf 73}, 064030 (2006)
  [gr-qc/0512160].

\bibitem{Berti:2007zu} 
  E.~Berti, J.~Cardoso, V.~Cardoso and M.~Cavaglia,
  Phys.\ Rev.\ D {\bf 76}, 104044 (2007)
  [arXiv:0707.1202 [gr-qc]].

\bibitem{Dominik:2012} 
  M.~Dominik, K.~Belczynski, C.~Fryer, D.~E.~Holz, E.~Berti, T.~Bulik, I.~Mandel and R.~O'Shaughnessy,
  Astrophys.\ J.\  {\bf 759}, 52 (2012)
  [arXiv:1202.4901 [astro-ph.HE]].

\bibitem{Dominik:2013} 
  M.~Dominik, K.~Belczynski, C.~Fryer, D.~E.~Holz, E.~Berti, T.~Bulik, I.~Mandel and R.~O'Shaughnessy,
  Astrophys.\ J.\  {\bf 779}, 72 (2013)
  [arXiv:1308.1546 [astro-ph.HE]].

\bibitem{Dominik:2014} 
  M.~Dominik {\it et al.},
  Astrophys.\ J.\  {\bf 806}, 263 (2015)
  [arXiv:1405.7016 [astro-ph.HE]].

\bibitem{de Mink:2016}
  I.~Mandel and S.~E.~de Mink,
  arXiv:1601.00007 [astro-ph.HE].

\bibitem{Marchant:2016} 	
  P.~Marchant, N.~Langer, P.~Podsiadlowski, T.~Tauris and T.~Moriya,
  arXiv:1601.03718 [astro-ph.SR].

\bibitem{Punturo:2010zz} 
  M.~Punturo {\it et al.},
  Class.\ Quant.\ Grav.\  {\bf 27}, 194002 (2010).

\bibitem{Barausse:2011vx} 
  E.~Barausse, V.~Cardoso and G.~Khanna,
  Phys.\ Rev.\ D {\bf 84}, 104006 (2011)
  [arXiv:1106.1692 [gr-qc]].

\bibitem{Seoane:2013qna} 
  P.~A.~Seoane {\it et al.} [eLISA Collaboration],
  arXiv:1305.5720 [astro-ph.CO].

\bibitem{Seto:2001qf} 
  N.~Seto, S.~Kawamura and T.~Nakamura,
  Phys.\ Rev.\ Lett.\  {\bf 87}, 221103 (2001)
  [astro-ph/0108011].

\end{thebibliography}
\end{document}